\documentstyle[twocolumn,prl,aps,epsfig]{revtex}

\begin{document}
\title{Longitudinal magnon in the tetrahedral
        spin system
       $\rm Cu_2Te_2O_5Br_2$ near quantum criticality
         }

\author{C. Gros$^1$, P. Lemmens$^{2,4}$, M. Vojta$^3$,
        R. Valent\'\i$^1$, K.-Y. Choi$^4$,
        H. Kageyama$^5$, Z. Hiroi$^5$,
        N.V. Mushnikov$^5$, T. Goto$^5$
        M. Johnsson$^6$, P. Millet$^7$
}

\address{$^1$Fakult\"at 7, Theoretische Physik,
         Universit\"at des Saarlands,
         D-66041 Saarbr\"ucken, Germany.}

\address{$^2$ Max Planck Institute for Solid State Research,
D-70569 Stuttgart, Germany}

\address{$^3$ Theoretische Physik III, Universit\"at Augsburg,
         D-86135 Augsburg, Germany}

\address{$^4$ 2. Physikalisches Institut, RWTH Aachen,
         D-52056 Aachen, Germany}

\address{$^5$ Institute for Solid State Physics, Univ. of Tokyo
         Kashiwa-shi, Chiba 277-8581,
         Kashiwa, Japan}

\address{$^6$ Department of Inorganic Chemistry,
        Stockholm Univ., S-10691 Stockholm, Sweden}

\address{$^7$ Centre d'Elaboration de Mat\'eriaux
         et d'Etudes Structurales, CEMES/CNRS, F-31062 Toulouse, France}

\maketitle

\begin{abstract}
We present a comprehensive study of the coupled tetrahedra-compound $\rm
Cu_2Te_2O_5Br_2$ by theory and experiments in external magnetic fields. We
report the observation of a longitudinal magnon in Raman scattering in the
ordered state close to quantum criticality. We show that the excited
tetrahedral-singlet sets the energy scale for the magnetic ordering
temperature $T_N$. This energy is determined experimentally. The ordering
temperature $T_N$ has an inverse-log dependence on the coupling parameters
near quantum criticality.
\end{abstract}

\begin{figure}
(a)
 \centerline{\epsfig{figure=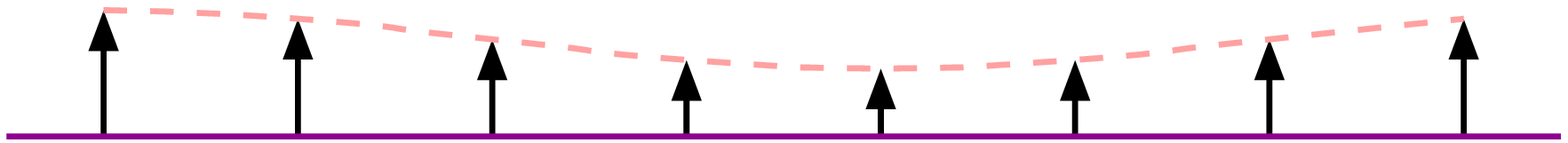,width=10cm}}
 \vspace{0.3cm}
\hspace{0.3cm}(b)
\centerline{\epsfig{figure=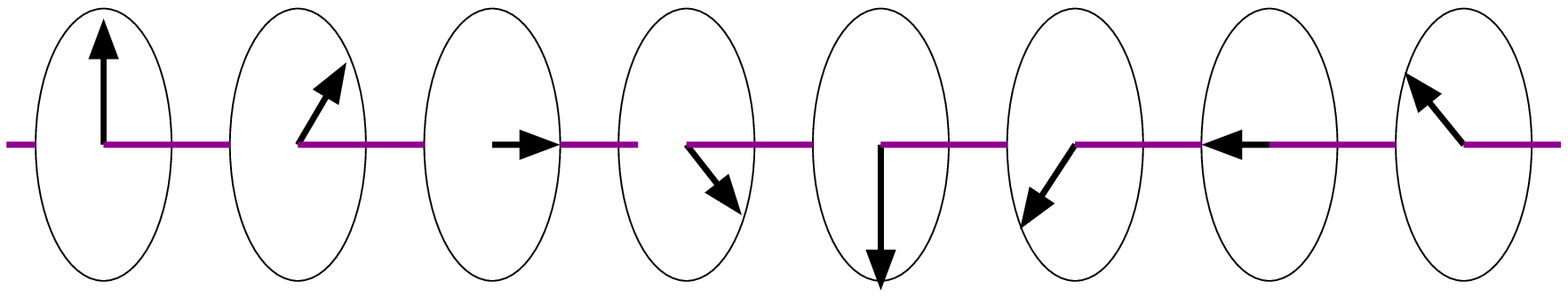,width=10cm}}
\caption{Schematic representation of (a) a longitudinal
magnon (b) a transversal magnon.
        }
\label{tm}
\end{figure}

\newpage\
{\bf Introduction}.- Quantum fluctuations in antiferromagnetic insulators
lead to a reduction of the magnetic moment and to a new mode in which the
magnitude of the local order parameter oscillates, the longitudinal magnon
(LM)\cite{Affleck} (see Fig.\ \ref{tm}(a)). This elementary excitation is
absent in classical magnets, where the excitations perform
a precession of the moment around its equilibrium position
and are therefore transversaly polarized (see Fig.\ \ref{tm}(b)).
The longitudinal mode is difficult to observe since it
occurs only in quantum spin systems with reduced dimensionality. Only
recently a LM has been detected by inelastic neutron scattering in quantum
spin-$\frac{1}{2}$ \cite{lake00,zheludev02} and spin-1 \cite{raymond99,enderle99}
chain compounds.

Quantum spin-fluctuations are of special importance in quasi-zero dimensional
systems with weakly coupled spin clusters. These lattices allow for quantum phase
transitions between magnetically ordered states and non-magnetic phases with a spin
gap. The recently discovered \cite{johnsson00} spin-tetrahedral compounds $\rm
Cu_2Te_2O_5X_2$ (X=Cl, Br) have been shown to order at transition temperatures
$T_N^{(\rm Cl)}=18.2~{\rm K}$ and $T_N^{(\rm Br)}=11.4~{\rm K}$ which are strongly
suppressed with respect to the magnitude of the intra-tetrahedral couplings
\cite{lemmens02}. Unconventional Raman-scattering has been found in the magnetic
channel \cite{lemmens02,brenig01} and the occurrence of low-lying singlet
excitations has been proposed \cite{johnsson00}.
Plateaus in the magnetization have been predicted
for a related linear chain of spin tetrahedra \cite{totsuka}.

The nature of the ordered states in $\rm Cu_2Te_2O_5X_2$ has
not yet been settled. The ordering temperature
$T_N^{(\rm Cl)}$ and $T_N^{(\rm Br)}$
decreases and rises respectively with an external magnetic field
\cite{lemmens02}. This unusual magnetic field-induced stabilization
of $T_N^{(\rm Br)}$ motivated in part the present study.
A decrease of $T_N$ is typical for an antiferromagnet
in the classical limit. We will show later on that $T_N$
may rise near a quantum-phase
transition. This results then indicates
$\rm Cu_2Te_2O_5Br_2$ to be close to criticality.

Here we report the observation of a LM in
$\rm Cu_2Te_2O_5Br_2$ by Raman scattering
in a magnetic field and present the evolution of
this mode under the influence of an
external magnetic field. We believe that this study constitutes
 the first time that a longitudinal magnon
is detected optically as well as the first observation of
such a mode in a tetrahedral spin system, i.e. in a system with
 an even number of spins per unit cell.    Furthermore, high-field magnetization
and other thermodynamic data on pure and substituted
$\rm Cu_2Te_2O_5(Br_xCl_{1-x})_2$ are
compared via a mean-field analysis which allows
to determine the microscopic parameters for $\rm Cu_2Te_2O_5Br_2$.
We find, interestingly, that the scale of the
ordering temperature $T_N$ is set by the
(non-magnetic) excited singlet of the copper tetrahedron and that
$T_N$ has an essential singularity at criticality.

\begin{figure}
 \centerline{\epsfig{figure=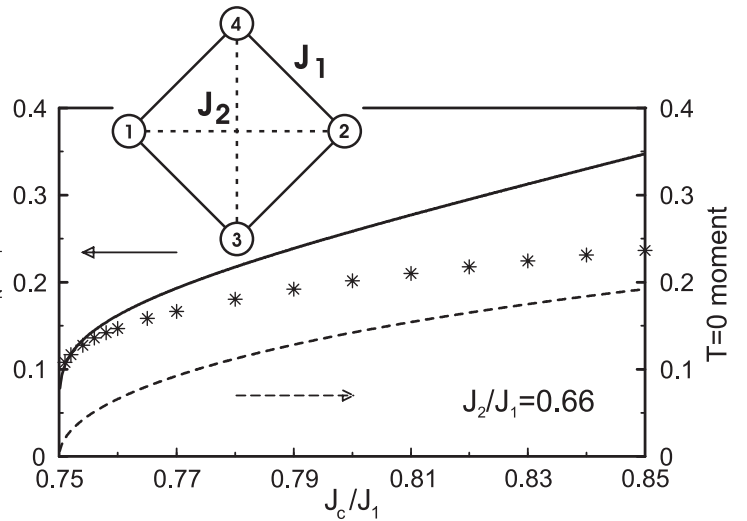,width=10cm}}
 \vspace{0.5cm}
\caption{Ne\'el-temperature (left scale) and spin tetrahedra mean-field
moment (right scale) for $J_2/J_1=0.66$. Shown is the analytic approximation
to leading order by Eq.\ (\ref{eq_T_N}) (solid line) and the numerical
solution of the self-consistency equation (\ref{eq_self}) (stars). The
dashed line (right scale) is the magnetic moment at zero-temperature Eq.\
(\ref{eq_M_0}). Note the logarithmic singularity in $T_N$ for
$J_c\to0.75J_1$. Inset: Cu tetrahedron with exchange couplings $J_1$ and
$J_2$ (solid/dashed lines).
        }
\label{fig_MF_TN}
\end{figure}

{\bf Mean-field approach}.- We assume that the basic spin-cluster in this compound
is given by the copper tetrahedron (see inset in Fig.\ (\ref{fig_MF_TN})). We denote
with $s_{kl}$ the spin-singlet state of two intra-tetrahedral sites $k$ and $l$ and
with $t_{kl}^\alpha$ the respective triplet states, with $\alpha=\pm1,0$. We start
by considering the eigenstates of the isolated tetrahedra
 with $H_{t}= J_1 [({\bf S}_1+ {\bf S}_2) \cdot
 ({\bf S}_3+ {\bf S}_4)] + J_2 ({\bf S}_1 \cdot
 {\bf S}_2 + {\bf S}_3 \cdot
 {\bf S}_4)$, which consist of two singlets,
three triplets and one quintuplet.

For $J_2<J_1$ the ground state singlet $\psi_{s1}$ and the
excited singlet $\psi_{s2}$ are
\begin{equation}
\psi_{s1} = {-1\over\sqrt3}
\left[ t_{12}^0 t_{34}^0 - t_{12}^+ t_{34}^- - t_{12}^- t_{34}^+
\right]~,\qquad
\psi_{s2} = s_{12} s_{34}
\label{singlet}
\end{equation}
with eigenenergies
$E_{s1} = -2J_1+J_2/2$ and $E_{s2} = -3J_2/2 = E_{s1} + \Delta E_{s2}$, with $\Delta
E_{s2}=2J_1-2J_2$. The three triplets $\psi_{t1}^\alpha$,
$\psi_{t2}^\alpha$ and $\psi_{t3}^\alpha$  have the (excitation) 
energies $\Delta E_{t1}=J_1$, $\Delta
E_{t2} = \Delta E_{t3} =2J_1-J_2$ with respective eigenstates
\begin{eqnarray}
\psi_{t1}^0 = {1\over\sqrt2}
(t_{12}^+ t_{34}^- - t_{12}^- t_{34}^+),\nonumber\\
\psi_{t1}^- = {1\over\sqrt2}
(t_{12}^0 t_{34}^- - t_{12}^- t_{34}^0) \nonumber\\
\psi_{t1}^+ = {1\over\sqrt2}
(t_{12}^+ t_{34}^0 - t_{12}^0 t_{34}^+)\nonumber\\
\psi_{t2}^\alpha = s_{12}^{\phantom{\alpha}}t_{34}^\alpha, \qquad
\psi_{t3}^\alpha = t_{12}^\alpha s_{34}^{\phantom{\alpha}.}
\label{tetra_states}
\end{eqnarray}
%
 The quintuplet has
the energy $\Delta E_{q} = 3J_1$. The inter-tetrahedra couplings can be described in
a mean-field approach by \cite{note_M}:
\begin{equation}
H_{MF} = -J_c M\left( S_1^z + S_2^z -S_3^z - S_4^z
              \right)~,\quad\mbox{and}\quad 
M= \frac{1}{4}\langle S_1^z + S_2^z -S_3^z - S_4^z\rangle
\label{eq_H_MF}
\end{equation}
with
M being the staggered
magnetization
order parameter.
$J_c$ is here the sum over all inter-tetrahedra
couplings.

The mean-field Hamiltonian $H_{MF}$ couples $\psi_{s1}$ and $\psi_{t1}^0$ leading to
new eigenstates for $H=H_t + H_{MF}$:
\begin{eqnarray}
|\varphi\rangle = \cos\varphi\, |\psi_{s1}\rangle
             + \sin\varphi\, |\psi_{t1}^0\rangle,\nonumber \\
|\tilde\varphi\rangle = \sin\varphi\, |\psi_{s1}\rangle
             - \cos\varphi\, |\psi_{t1}^0\rangle
\label{eq_phi}
\end{eqnarray}
with $\langle\tilde\varphi|\varphi\rangle=0$ and new energies
\begin{equation}
\Delta E_{\varphi,\tilde\varphi} =
{J_1\over2}\left[
1\mp\sqrt{1 + 32 M^2 J_c^2/(3J_1^2)
         } \right]
\label{eq_E_phi}
\end{equation}
with $\tan\varphi = -\Delta E_\varphi\sqrt{6}/(4J_cM)$.
  $|\varphi\rangle$ is the ground-state and
$|\tilde\varphi\rangle$ can be identified as a longitudinal magnon
excitation.
The physical interpretation of this excitation is as follows. When $J_c$=0
we have isolated tetrahedra and $|\tilde\varphi\rangle$ would correspond
to the excited intra-tetrahedral triplet state  $|\psi_{t1}^0\rangle$. For
$J_c$ $\neq$ 0,  $|\tilde\varphi\rangle$
 evolves continuously
from $\psi_{t_1}$ as a function of the inter-tetrahedral
coupling $J_c$ and becomes soft at the transition-point
to the ordered state. 
The molecular field couples also $\psi_{t1}$ with the quintuplet $\psi_q$,
though we neglect this coupling here since we are interested in phases
with low transition temperatures $T_N$ for which the
high-energy quintuplet does not contribute significantly.

The calculation of the staggered magnetization
 $M= Tr[(S_1^z + S_2^z -S_3^z - S_4^z){\rm e}^{-\beta
 H}]/(4Z)$
 (Eq. \ref{eq_H_MF})
leads to the following 
 self-consistency equation,
\begin{equation}
M = {
    {\rm e}^{-\beta E_\varphi}
   -{\rm e}^{-\beta E_{\tilde\varphi}}
    \over Z}
    \sqrt{2\over 3}
          {\tan\varphi \over 1+\tan^2\varphi}~,
\label{eq_self}
\end{equation}
where  $\beta=1/T$
and $Z$ is the partition function for the coupled tetrahedra system,
i.e.
$Z = {\rm e}^{-\beta
 E_{s1}} + {\rm e}^{-\beta
 E_{s2}} + ...~$ . For $J_c=
J_c^{(qc)}=3J_1/4$ the magnetization M goes to zero
 and the system shows a second-order phase transition at $T_N$.


{\bf Results}.-  The transition-temperature $T_N$ can be obtained from
(\ref{eq_self}) by imposing $M=0$. Assuming that (i) $s_2$ is the lowest excited state of
a tetrahedron and (ii) at small temperatures only the leading order in a
$1/T$ expansion is contributing in Eq.\ (\ref{eq_self}),  T$_N$ can be 
analytically derived:  
\begin{equation}
T_N \simeq \Delta E_{s2}\,
\log^{-1}\left( J_c^{(qc)}/(J_c-J_c^{(qc)}) \right)~,
\label{eq_T_N}
\end{equation}
 $T_N$ shows  an inverse-log singularity close to the quantum critical
 point at $J_c = J_c^{(qc)}$. The critical
$J_c^{(qc)}=3J_1/4$ is independent of $J_2$. In Fig.\ (\ref{fig_MF_TN})
we plot $T_N$ as a function of $J_c$ both as obtained in the analytic
solution Eq.\ (\ref{eq_T_N}) and by solving numerically the
self-consistent equation Eq.\ \ref{eq_self}. Note that for the region $J_c \sim
J_c^{(qc)}$,  Eq.\ (\ref{eq_T_N}) provides a good approximation for $T_N$.
 
 The inverse-log dependence of the
Ne\'el-temperature implies that $T_N$ is substantial even near the
quantum critical point, as illustrated in Fig.\ (\ref{fig_MF_TN}), in
contrast to the magnitude of the zero-temperature magnetic moment,
\begin{equation}
M(T=0) = {1\over\sqrt6}
\sqrt{1-\left(J_c^{(qc)}/J_c\right)^2}~,
\label{eq_M_0}
\end{equation}
which has a standard mean-field form\cite{note_q} 
(compare  Fig.\ (\ref{fig_MF_TN})). For
$J_2>J_1$ the tetrahedral ground-state changes to $\psi_{s_2}$
and the non-magnetic singlet $\psi_{s_2}$ sets therefore
the scale for $T_N$.


{\bf States in an external field}.- An external longitudinal magnetic field
does not induce additional couplings in between the different eigenstates
but it leads to shifts in the respective eigenenergies. A transversal
magnetic field $B_x$ induces, on the other hand,   a coupling in between
$\psi_{t1}^0$ and $\psi_{t1}^{\pm}$ (see Eq.\ (\ref{tetra_states})). 
The mean-field ground state, which breaks rotational 
invariance, can  be written, in lowest order in $B_x$, as
\begin{equation}
|\varphi,B_x\rangle = \cos\alpha\, |\varphi\rangle
+ {\sin\alpha\over\sqrt 2}
\Big[ |\psi_{t1}^+\rangle + |\psi_{t1}^-\rangle
\Big],
\label{eq_phi_B_x}
\end{equation}
with $\tan\alpha= B_x\sin\varphi/(J_1-\Delta E_\varphi)$. In lowest order
in $B_x$, the ground-state energy,
$\Delta E_{\varphi,B_x}=E_{\varphi,B_x}-E_{\varphi}$,
\begin{equation}
\Delta E_{\varphi,B_x} = -B_x^2\sin^2\varphi/(J_1-\Delta E_{\varphi}),
\label{B_dependence}
\end{equation}
decreases quadratically with $B_x$. 

This result has an interesting consequence for the transition temperature.
The energy of the excited singlet $E_{s2}$ is not affected by $B_x$, its
relative energy to the ground state $\Delta E_{s2}$ increases consequently
with $B_x$ (compare Eq.\ (\ref{B_dependence})). 
 Eq.\ (\ref{eq_T_N}) tells us then that the N\'eel temperature
also increases    with $B_x$ \cite{note1}.
An order-of magnitude estimate of the effect for $\rm
Cu_2Te_2O_5Br_2$ and $B_x=13\,{\rm T}$ yields $\Delta T_N\approx 0.8\,{\rm
K}=0.56\,{\rm cm}^{-1}$ which agrees well with the experimentally 
observed raise of $\sim1\,{\rm K}$ reported already in Ref.\ \cite{lemmens02}.

For the longitudinal magnon, a calculation analogous to
Eq.\ (\ref{B_dependence}) leads to
\begin{equation}
\Delta E_{\tilde\varphi,B_x} = -B_x^2\cos^2\varphi/(J_1-\Delta E_{\tilde\varphi}),
\label{deltaB}
\end{equation}
The resulting change $\Delta E_{\tilde\varphi,B_x} = E_{\tilde\varphi,B_x} 
-E_{\tilde\varphi}$ in the longitudinal-magnon energy
is positive, as $J_1-\Delta E_{\tilde\varphi}<0$, and substantially
larger than the shift for the ground-state
$\Delta E_{\varphi,B_x}$ (and correpondingly for $E_{s2}$) since
$\cos^2\varphi\gg\sin^2\varphi$. As we shall discuss in the next
sections,
 this trend is qualitatively in
agreement with the Raman data presented below.

\begin{figure}
 \centerline{\epsfig{figure=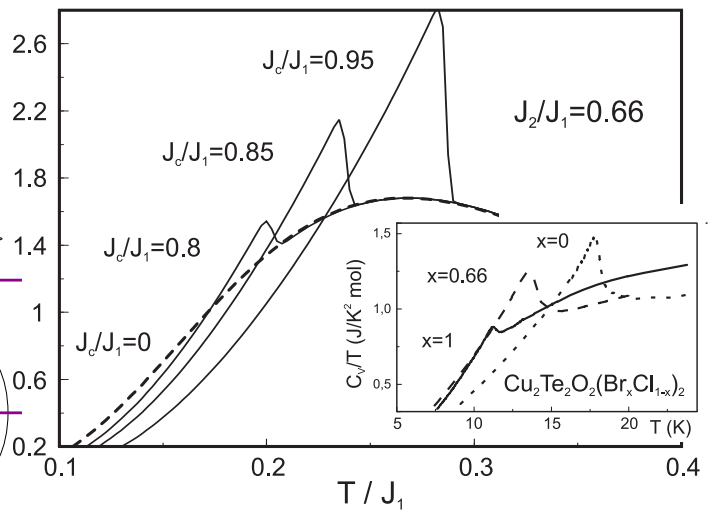,width=10cm}}
 \vspace{1cm}
\caption{Mean-field results for the specific heat of spin tetrahedra coupled
by $J_c$. The inset shows the specific heat of $\rm
Cu_2Te_2O_5(Br_{x}Cl_{1-x})_2$ with x=1, 0.66 and 0. The data with x=0 and 1
are compiled from \protect\cite{lemmens02}.
        }
\label{fig_MF_cV}
\end{figure}

{\bf Substitution experiments}.- Specific heat and high-field magnetization
have been measured on $\rm Cu_2Te_2O_5(Br_xCl_{1-x})_2$ powder samples with
$x=1,0.75,0.66$ and $0.0$ \cite{johnsson00,lemmens02}. Substituting Br by Cl
leads to a continuous decrease of the unit cell volume by 7\% (x=0) and an
increase of the transition temperature from 11.4 to 18.2~K. Also other
physical properties change continuously with substitution
\cite{lemmens-kashiwa}.  In our coupled-cluster model we expect that the
decrease of the unit cell volume is described by an increase of the coupling
constant $J_c$. We have calculated the specific heat 
\begin{equation}
C_v = \beta^2 \langle (H - \langle H \rangle)^2 \rangle
\end{equation}
in the mean-field approximation. In Fig.\ \ref{fig_MF_cV} the results
are shown for various $J_c$ values. In the inset of Fig.\ \ref{fig_MF_cV} the evolution
of the experimentally determined specific heat as a function
of substitution $x$ is presented. Note that the mean-field results
for increasing $J_c$
reproduce
the continuous shift of the specific heat anomaly to higher temperatures
wit decreasing x.

\begin{figure}
 \centerline{\epsfig{figure=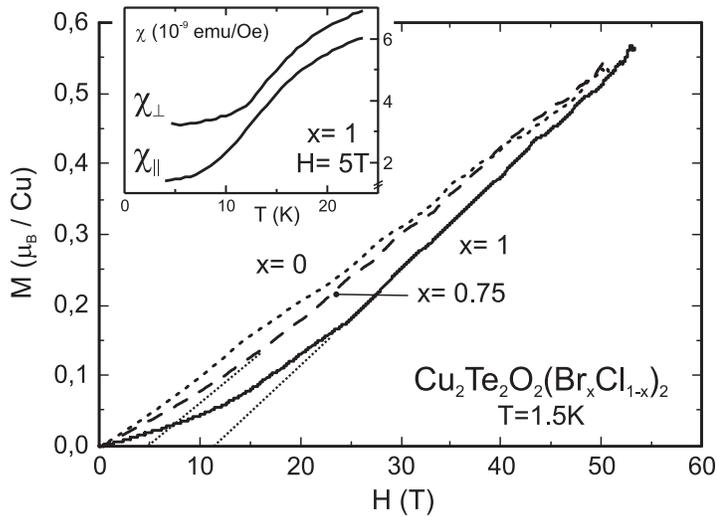,width=10cm}}
 \vspace{0.5cm}
\caption{High-field magnetization of $\rm Cu_2Te_2O_5(Br_{x}Cl_{1-x})_2$
powder samples for x=1, 0.75 and 0. The data with x=0.66 is omitted here for
clarity. Dotted lines correspond to high field extra\-polations. The inset
shows the anisotropic magnetic susceptibility of $\rm Cu_2Te_2O_5Br_2$
single crystals. } \label{high_field}
\end{figure}

 A further support to the interpretation of these systems as that of
 coupled tetrahedra with a mean-field $J_c$ inter-tetrahedra coupling
 and 
with a $T_N$ ordering temperature, 
 is obtained from high-field magnetization measurements which are presented in Fig.\
 \ref{high_field}.
 We observe
a finite slope  for
all samples at small fields which increases with decreasing x. Extrapolating M(H)
from high to low fields a finite crossover field, $\rm H_{cross}$= 12~T, 5~T
and $\approx$~0~T respectively, is defined as a function of decreasing x. No direct
evidence for a plateau in the magnetization is found.  A plateau at M=1/2 is
predicted for a 1D chain of spin tetrahedra with parameters placing the
system in the gapped phase \cite{totsuka}, what leads us to conclude
that these systems cannot be described as chains of tetrahedra.  The finite slope in M(H) at small
fields and even for x=1 is intrinsic and points to the underlying weak
N\'eel state. The corresponding anisotropy is observed in single crystal
susceptibility shown in the inset of Fig.\ \ref{high_field}. The transition
is evident as a kink in $\rm \chi(T)$ with a magnetic field perpendicular to
the c-axis.

\begin{figure}
 \centerline{\epsfig{figure=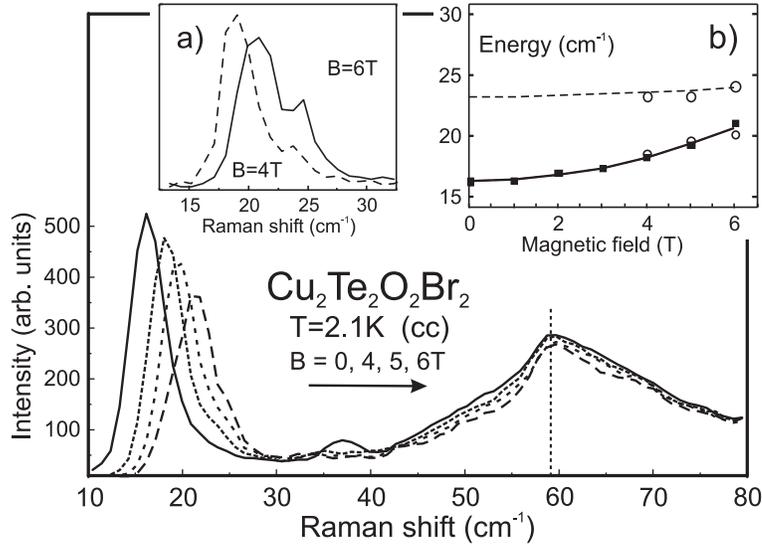,width=10cm}}
 \vspace{1cm}
\caption{Raman spectra of $\rm Cu_2Te_2O_5Br_2$ in a magnetic field. The
insets show a) spectra with higher resolution and b) the shift of the $\nu_{sing}=23.2\,{\rm
cm}^{-1}$ mode (upper open symbols) and the $\nu_{long}=16.3\,{\rm
cm}^{-1}$ mode (lower full and open symbols) 
as a function of the magnetic field. The dashed  line shows the field
dependence of the transition temperature \protect\cite{lemmens02}
 and the full line is a fit to the data proportional to the square
 of the magnetic field.  The
open (full) symbols show data with high resolution (normal resolution).
        }
\label{fig_raman}
\end{figure}

{\bf Raman spectroscopy}.- We have performed Raman scattering experiments on
c-axis oriented single crystals with diameters $\rm \O \approx 0.2~mm$ and
length $\rm l \approx$~1~mm. The used scattering geometry in (cc) light
scattering polarizations corresponds to \textbf{A} symmetry
\cite{lemmens02}. The magnetic field has been applied perpendicular to the
light scattering polarization. In Fig.\ (\ref{fig_raman}) we present Raman
data for $\rm Cu_2Te_2O_5Br_2$ in magnetic fields up to $\rm 6\, T$. We
observe a shift of the low-energy magnetic mode at $\nu_{long}=16.3\,{\rm
cm}^{-1}=23.4\,{\rm K}$ (for $B = 0$) to higher energies as a function of B
and the appearance of an
additional, magnetic field-induced mode at $\nu_{sing}=23.2\,{\rm
cm}^{-1}=33.2\,{\rm K}$ ($B \neq 0$). In the inset Fig.\ (\ref{fig_raman}a) the low
energy spectrum is displayed with smaller optical slit width and higher
spectral resolution. The intensity of the higher energy mode increases with
field. It is not observable for $B = 0$.

In Fig.\ (\ref{fig_raman}b) the energies of the respective modes are shown as a
function of the magnetic field. While the higher energy mode does not show an
appreciable magnetic field dependence (upper open symbols),
 the lower energy mode shows a nonlinear
dependence on the magnetic field (lower full and open symbols).
 The full line in Fig.\ (\ref{fig_raman}b) is a fit
to the data proportional to the square of the magnetic field. The dashed curve
describing the higher energy mode is proportional to the weaker, positive field
dependence of the transition temperature as determined by specific heat and magnetic
susceptibility measurements \cite{lemmens02}.
In the following we shall argue that these two modes at  $\nu_{long}=16.3\,{\rm
cm}^{-1}$ and at  $\nu_{sing}=23.2\,{\rm cm}^{-1}$ can be identified
as a longitudinal magnon mode and  as an  excitation to the second
singlet $\psi_{s2}$ in Eq.\ (\ref{singlet}) respectively.


{\bf Excited singlet}.- We interpret the additional higher energy mode at
$\nu_{sing}=23.2\,{\rm cm}^{-1}$ presented in Fig.\ (\ref{fig_raman}) as a
transition to the second singlet $\psi_{s2}=s_{12}s_{34}$. Since
this system is noncentrosymmetric, there is a nonzero
Dzyaloshinski-Moriya (DM)\cite{DM}
interaction.  Assuming a DM
 contribution to the Raman operator, i.e.
 $H_R^{(DM)}\sim {\bf D}_{ij}\cdot({\bf
S}_i\times {\bf S_j})$ we find a non-zero Raman matrix-element
\[
\langle \psi_{s2}|H_R^{(DM)}|\varphi,B^x\rangle
\sim\sin\alpha\sim B^xM~.
\]
The $23.2\,{\rm cm}^{-1}$-mode would therefore be observable only in the
ordered phase and in an external magnetic field, consistent with experiment.
Having identified this mode with the transition to $\psi_{s2}$ we have then
that
\begin{equation}
\Delta E_{s2}=2J_1-2J_2 = \nu_{sing}\sim 33\,{\rm K}~.
\label{eq_E_s2}
\end{equation}
Considering (\ref{eq_E_s2}) and the magnetic susceptibility
\cite{johnsson00,lemmens02} for $T>T_N$ we find a good fit with
$J_1 \sim 47\,{\rm  K}$ and $J_2 \sim 31\,{\rm  K}$ which yields
$J_2/J_1 \sim 0.66$.
The experimental transition temperature \cite{lemmens02} of $T_N^{(\rm
Br)}=11.4~{\rm K}$ for $\rm Cu_2Te_2O_5Br_2$ implies via the
self-consistency condition (\ref{eq_self}) that $J_c \sim 0.85J_1$.

Moreover, recalling Eqs.\ (\ref{eq_T_N}) and (\ref{B_dependence}),
for $B_x=6\,{\rm T}$ the energy $\Delta E_{s2}$ of the excited singlet
shifts by about $0.12\,{\rm cm}^{-1}$. This increase
is too small to be resolved by Raman, although the data presented in
inset b) of Fig.\ \ref{fig_raman} seem to indicate a small increase.


{\bf Longitudinal magnon}.- We observe that the mean-field Hamiltonian
(\ref{eq_H_MF}) leads to a $\bf Q=0$ ordering for $J_c>0$ and the soft
longitudinal magnon $|\tilde\varphi\rangle$ should be directly observable
with Raman. For $J_c<0$ ordering with ${\bf Q}=\pi$ would occur and
additional backfolding to the zone-center via residual lattice distortions
would be necessary. The matrix-element
$\langle\tilde\varphi|H_R|\varphi\rangle$ of the Raman-operator $H_R\sim
{\bf S}_i\cdot {\bf S}_j$ ($i,j=1,\dots,4$) is $\sim\cos\varphi\sin\varphi$.
It vanishes in the decoupled-tetrahedra limit $J_c=0$, $\varphi=0$ and the transition
should be observable only in the ordered phase.

Summarizing, the Raman mode at $\nu_{long}=$$16.3\,{\rm cm}^{-1}=23.4\,{\rm K}$ in Fig.\
\ref{fig_raman} (i) has been shown \cite{lemmens02} to become soft at the
ordering temperature, (ii) it is observable only in the condensed phase and its
energy increases quadratically (compare Eq.\ (\ref{deltaB})) with the
field. We interpret it therefore as a longitudinal magnon.

The energy of this mode is strongly suppressed below its mean-field energy
$E_{\tilde\varphi}- E_{\varphi}=54\,{\rm K}$ by dispersion. We
can estimate the magnitude of this suppression by comparison with the
results of a bond-operator-theory for a coupled dimer system
\cite{sommer01,normand97} (alternatively one may use a
generalized RPA-approach \cite{jensen}).
The effective dimer-states are $(\psi_{s1},\psi_{t1}^\alpha)$.
The gap of the longitudinal magnon has the
form \cite{sommer01}
\begin{equation}
\Delta_{\rm long} = \Delta_{\rm max}
\sqrt{1-\left(J_c^{(qc)}/J_c\right)^2}~.
\label{eq_gap_long}
\end{equation}
For $\rm Cu_2Te_2O_5Br_2$ we have $J_c=0.85J_1$ ($(J_c^{(qc)}/J_c)^2=0.78$) and
$\Delta_{\rm long}\approx 0.47\Delta_{\rm max}$. 
The energy scale $\Delta_{\rm max}$ occuring in Eq.\ (\ref{eq_gap_long})
is set by the longitudinal magnon-gap in the classical N\'eel ordered state,
i.e.\ in the limit of strong inter-dimer (tetrahedra) couplings, where
it has the value $\Delta_{\rm max}\to J_c$. 
 With $J_c\approx 0.85J_1$ and $J_1\approx47\,{\rm K}$ we
then find for $\rm Cu_2Te_2O_5Br_2$ that
 $\Delta_{\rm long}\approx 19\,{\rm K}$, which is
qualitatively in agreement with the experimental value $\Delta_{\rm long}^{(\rm
exp)} = \nu_{long} = 23.4\,{\rm K}$, see the inset Fig.\ \ref{fig_raman}
b.
Due to the renormalization of $\Delta_{\rm long}$ by the
fluctuations near the quantum-critical point, see Eq.\ (\ref{eq_gap_long}),
we cannot easily quantitatively estimate its dependence on an external magnetic
field, as presented in inset b) of Fig.\ \ref{fig_raman}.


{\bf Comparison with ${\bf \rm Cu_2Te_2O_5Cl_2}$}.- The magnitude of the
intra-tetrahedral parameters have been estimated \cite{johnsson00} from
susceptibility measurements to be similar both for $\rm Cu_2Te_2O_5Br_2$ and the
isostructural $\rm Cu_2Te_2O_5Cl_2$. The substantially enhanced N\'eel temperature
\cite{lemmens02} of $T_N^{(\rm Cl)}=18.2~{\rm K}$ indicates a larger inter-dimer
coupling for the Cl-compound. This notion is consistent with the Raman-results for
doped compounds which indicates a harding of the excitations with Cl-content, as
predicted by Eq.\ (\ref{eq_gap_long}) \cite{lemmens-kashiwa}.
We did not attempt a quantitative analysis of the
coupling parameters for $\rm Cu_2Te_2O_5Cl_2$ since we
were not able to observe the second singlet as in $\rm Cu_2Te_2O_5Br_2$.
Indeed, a generalized tight-binding analysis of
band-structure calculations\cite{valenti03} indicates that the
ratio of the intra-dimer couplings $J_2/J_1$ in $\rm Cu_2Te_2O_5Cl_2$ is
smaller than in $\rm Cu_2Te_2O_5Br_2$. These findings also suggest that for $\rm
Cu_2Te_2O_5Cl_2$ the excited singlet with energy $E_{s2}=2J_1-2J_2$
should 
probably be
located in the energy-range of the magnetic continuum and thus not observable
separately.


{\bf Conclusions}.- We have presented a comprehensive set of theory and experimental
data indicating that the isostructural spin-tetrahedral compounds $\rm
Cu_2Te_2O_5(Br_xCl_{1-x})_2$ constitute a series of systems with a systematic
variation of the microscopic parameters with respect to a quantum critical
transition. We have pointed out the importance of the low-lying singlet for the
magnetic state and reported the observation of a low-energy longitudinal magnon.

We acknowledge the support of the German Science Foundation (DFG SP1073,
SFB484), and important discussions with Wolfram Brenig, Frederic Mila,
Jens Jensen and T. Saha-Dasgupta.

\begin{thebibliography}{99}
\bibitem{note_q}
For $J_c/J_1\to\infty$ the coupling to
the quintuplet $\psi_q$ neglected in
Eq.\ (\ref{eq_M_0})
would contribute and the
zero-$T$ moment would take the mean-field value
of $1/2$.
\bibitem{note1}  Note that there is a smaller decreasing
                contribution in $B_x$  
		to the N\'eel temperature coming
		from the log-term in Eq.\ (\ref{eq_T_N})
                due to an enhancement of the effective
                $J_1$.
\end{thebibliography}
\end{document}